\documentclass[reprint,
superscriptaddress,
 amsmath,amssymb, aps, pra,
]{revtex4-2}
\usepackage[colorlinks, citecolor = magenta]{hyperref}

\usepackage{graphicx}
\usepackage{dcolumn}
\usepackage{bm}
\usepackage{braket}
\usepackage{natbib}

\begin{document}


\title{Hyperentanglement-enhanced quantum illumination}
\author{Ashwith Varadaraj Prabhu}
\affiliation{Indian Institute of Science, Bengaluru - 560012, India}
\author{Baladitya Suri}
\affiliation{Dept. of Instrumentation \& Applied Physics, Indian Institute of Science, Bengaluru - 560012, India}
\author{C. M. Chandrashekar}
\affiliation{Dept. of Instrumentation \& Applied Physics, Indian Institute of Science, Bengaluru - 560012, India}
\affiliation{The Institute of Mathematical Sciences,  Chennai - 600113, India}


\begin{abstract}
 \noindent In quantum illumination, the signal mode of light, entangled with an idler mode, is dispatched towards a suspected object bathed in thermal noise and the returning mode, along with the stored idler mode, is measured to determine the presence or absence of the object. In this process, entanglement is destroyed but its benefits in the form of classical correlations and enlarged Hilbert space survive. Here, we propose the use of  probe state \textit{hyperentangled} in two degrees of freedom -- polarization and frequency, to achieve a significant 12dB performance improvement in error probability exponent over the best known quantum illumination procedure in the low noise regime. We present a simple receiver model using four optical parametric amplifiers (OPA) that exploits hyperentanglement in the probe state to match the performance of the feed-forward sum-frequency generator (FF-SFG) receiver in the high noise regime.  By replacing each OPA in the proposed model with a FF-SFG receiver, further 3dB improvement in the performance of a lone FF-SFG receiver can be seen.
\end{abstract}


\maketitle

\section{Introduction}
 Detecting the presence or absence of an object in real space using low intensity probe states is often a challenging task due to the surrounding thermal noise. In such an object-detection scenario, protocols using quantum states as probes have been developed, where a probe state is transmitted towards the suspected object and the returning state is measured. The efficacy of different transmitter-receiver models is quantified using the probability of error in detecting the presence or absence of the suspected object\,\cite{pirandola2018advances, shapiro2020quantum}. For a given transmitter, the minimum error probability for deducing the object to be present is achieved when the measurement projects onto the positive eigenspace of $\rho_1-\rho_0$ where $\rho_0$ is the density operator of the returning state in absence of any object while $\rho_1$ is the density operator in its presence\,\cite{helstrom1976quantum}. The calculation of this minimum error probability is, in general, prohibitively difficult. The relatively tractable quantum Chernoff bound (QCB) provides an asymptotically tight bound on the error probability($p_e$)\,\cite{audenaert2007discriminating}. It is defined as $Q_{QCB}\equiv\underset{0\leq s \leq 1}{\text{min}}\text{Tr}\big[\rho_1^s\rho_0^{1-s}\big]$. After $N$ detection events, the bound on the error probability is given by,
\begin{align}
p_e^{(N)}\leq\frac{1}{2}\left(Q_{QCB}\right)^{N}\label{eqn:0}
\end{align}
 Several candidate probe states -- \textit{viz} coherent states, single photons\,\cite{lloyd2008enhanced}, entangled biphotons, two mode squeezed vacuum (TMSV) state have been considered for object-detection\,\cite{tan2008quantum}. Quantum Illumination (QI), in general, refers to the use of entangled states, where photons in one of the entangled states, treated as signal photons, are sent towards the suspected object, and the so-called idler photons in the other entangled state are kept back. Joint measurement of the returning photons and idler photons has been shown to be more effective than the use of classical states of light in detecting the presence of absence of the object in certain parameter regimes. With the mean number of thermal photons denoted by $N_B$ and reflectance of object by $\kappa$, the quantum Chernoff bound for QI using an entangled probe state such as TMSV, having $N_S$ mean number of photons in each entangled mode  was computed\,\cite{tan2008quantum} and a  6dB (factor of 4) improvement in the exponent of the error bound over the best classical object detection strategy was shown in the high noise and low signal intensity regime ($N_S\ll 1, \kappa \ll 1$ and $N_B\gg 1$)\,\cite{tan2008quantum, fnote}. After $N$ iterations, this bound is given by\,\cite{tan2008quantum}, 
\begin{align}
p_{e,QI}^{(N)}\leq e^{-\kappa NN_S/N_B}/2.
\end{align}
 The low noise regime, on the other hand, is characterised by the mean thermal photon number per temporal mode ($N_B$) being significantly less than unity ($N_B \ll 1$). In a typical QI protocol in this regime, a transmitter dispatches $N$ iterations of signal photons  spanning over $M$ temporal modes ($M\gg1$) towards a possible target of very poor reflectance $\kappa$. In the so-called ``bad regime"\,\cite{lloyd2008enhanced} characterized by $MN_B\ll 1$ and $\kappa \ll N_B/M$ within the low noise regime, the bound on the error probability is given by\,\citep{shapiro2009quantum},
\begin{align}
p_{e,QI}^{(N)}\leq e^{-NM\kappa^2/8N_B}/2,\label{eq:1}
\end{align} 
which is lower by a factor of $M$ in the exponent than a detection protocol employing only a single photon probe (SP)\,\cite{shapiro2009quantum},
\begin{align}
p_{e,SP}^{(N)}\leq e^{-N\kappa^2/8N_B}/2\,.
\end{align} 
 However, it should be noted that the use of a coherent state as probe (CS), which has associated error probability bound of $e^{-\kappa N_S}/2$ outperforms QI in this regime\,\cite{shapiro2009quantum}.  At the heart of it, the improved performance of QI stems from the larger Hilbert space of the entangled signal-idler system\,\cite{sacchi2005optimal}. Therefore, use of an enlarged Hilbert space through ``Hyperentanglement'' i.e. entanglement in more than one degree of freedom of the probe, without increasing the number of photons\,\cite{kwiat1997hyper} along with an appropriately  designed receiver can further increase the efficiency of QI. With various experiments demonstrating the generation of hyperentangled states of photon\,\cite{deng2017quantum}, its use for QI can be substantiated. While the enhancement effected by hyperentanglement in a continuous parameter estimation problem has been analysed in an earlier work\,\cite{walborn2018quantum}, this article shines light on the enhancement in what is essentially a hypothesis testing task. In this article, we propose object-detection using a probe state hyperentangled in polarization as well as frequency degrees of freedom\,\cite{chen2020recovering}. We show that the QCB for object-detection using hyperentangled photons gives a remarkable 12dB improvement in the exponent of the error probability over QI in the ``bad regime". For the receiver, we propose a setup that includes four optical parametric amplifiers (OPAs) that exploits the presence of correlations between the four pairs of modes for every iteration due to the hyperentanglement. This results in a 3dB improvement in high noise regime over the earlier proposals for practical quantum illumination receivers  using the optical parametric amplifier (OPA) and phase conjugate receiver (PCR)\,\cite{guha2009gaussian, zhang2015entanglement}. This performance matches the only other proposal that theoretically achieves the same using  sum-frequency generation receiver with feedforward (FF-SFG) receiver\,\cite{zhuang2017optimum}, whose implementation is, to our knowledge, more challenging. Finally, by replacing each OPA with a FF-SFG receiver, we also show that this hyperentanglement-enhanced FF-SFG receiver outperforms the lone FF-SFG receiver by 3dB in the error probability exponent. The next section details the 12dB performance improvement offered by hyperentanglement in the low noise regime and the two sections following the next analyse the performance of the proposed hyperentanglement-enhanced OPA receiver and the hyperentanglement-enhanced FF-SFG receiver in the high-noise regime. 

\begin{figure*}
\centering
\fbox{\includegraphics[width=11cm,height=7cm]{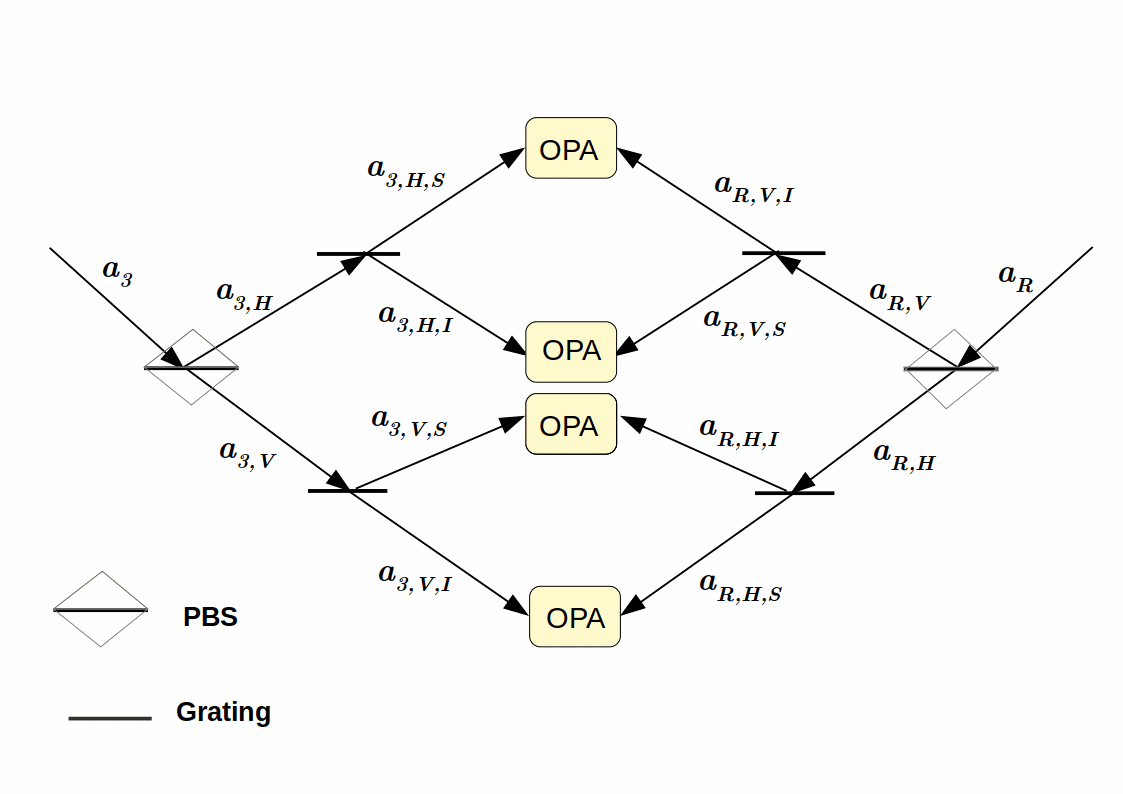}}
\caption{(Colour Online) Schematic representation of hyperentanglement enhanced OPA receiver. The stored spatial mode undergoes splitting by a polarizing beam splitter (PBS) and each of the two resulting modes are subjected to frequency dependent splitting using optical grating. The returning state goes through an identical procedure. The four pairs of correlated modes form the inputs to four OPAs.}
     \label{fig:2}
\end{figure*}

\begin{figure}
    \includegraphics[width=8.5cm]{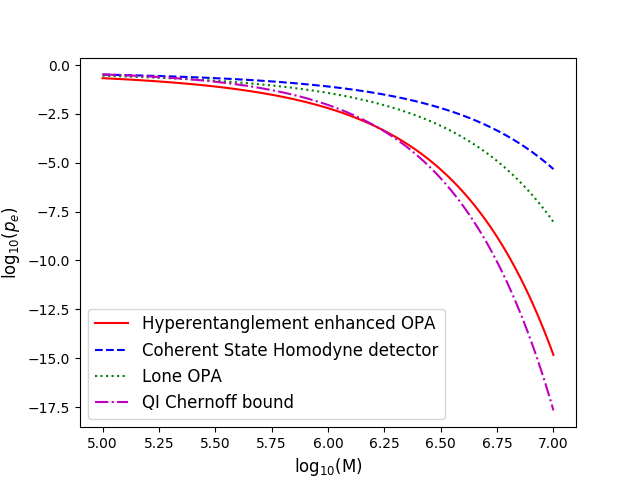}
    \caption{(Colour Online)Performance of different receivers.  The error probability of detection for hyperentanglement enhanced OPA receiver is close to the QCB for TMSV and is therefore better than the coherent state homodyne detector as well as lone OPA receiver in high noise regime. The parameters used for the plots are $N_S=\kappa=0.01, N_B=25$ and $G=1.005$.}
    \label{fig:1}
\end{figure}
    
\section{Hyperentanglement-enhanced sensitivity in the low noise regime}
 Ever since the first experimental demonstration of generation of hyperentangled photon pairs\,\cite{barreiro2005generation}, different procedures to generate hyperentanglement in different degrees of freedom  of photons have been successfully demonstrated. Here we will briefly outline one such procedure\,\cite{chen2020recovering} to generate states hyperentangled in polarization and frequency degrees of freedom. Using two identical type-II non-collinear SPDC generators driven by identical pumps, derived from a common pump, $\ket{\Phi}_1$ and $\ket{\Phi}_2$,  that are entangled in polarization degree of freedom\,\cite{kwiat1999ultrabright} are produced. For each of these states, the mean number of photons in each entangled mode of the state is $N_S^\prime$. The states have the Fock state represenation\,\cite{footnote},
 \begin{align}
\ket{\Phi}_1=\sum_{n=0}^\infty \frac{{N_S^\prime}^{n/2}}{\sqrt{2}(N_S^\prime+1)^{(n+1)/2}}&(\ket{n}_{1,H,S}\ket{n}_{1,V,I}\nonumber \\+\ket{n}_{1,H,I}\ket{n}_{1,V,S})\label{eq:pstate1}\\ 
\ket{\Phi}_2=\sum_{n=0}^\infty\frac{{N_S^\prime}^{n/2}}{\sqrt{2}(N_S^\prime+1)^{(n+1)/2}}&(\ket{n}_{2,H,S}\ket{n}_{2,V,I}\nonumber\\+\ket{n}_{2,H,I}\ket{n}_{2,V,S}).\label{eq:pstate2}
\end{align}
 Here, $S$ and $I$ refer to signal and idler frequency, $1$ and $2$ denote distinct spatial modes while $H$ and $V$ denote horizontal and vertical polarization respectively. The output spatial modes of a single non-collinear SPDC process are distinct but using appropriate lenses, the spatial overlap of the output modes can be increased such that the two spatial modes may be approximated to be identical. Thus, the signal and the idler frequency modes occupy the same spatial mode. If the original common pump had been used to drive a single SPDC generator, the mean number of photons in each of the two output modes would have been $N_S$ (say) and would be related to $N_S^\prime$ as $N_S^\prime=N_S/2$. The two spatial modes form the input to a balanced beam splitter to give,
\begin{align}
a_3=\frac{a_1+ia_2}{\sqrt{2}},\label{eq:beam1}\\
a_4=\frac{ia_1+a_2}{\sqrt{2}}.\label{eq:beam2}
\end{align}
The resulting state is entangled in frequency $\omega_S$ and $\omega_I$ besides polarization. A first order approximation in squeezing parameter of the resulting hyperentangled state and setting the relative pump phase to zero or $\pi$ yields a hyperentangled biphoton state\,\cite{chen2020recovering}. For the low noise regime, the time-bandwith product is adjusted such that the transmitter pulse spans over $M$ temporal modes to give the following output state\,\cite{chen2020recovering},
\begin{equation}
\begin{aligned}
\ket{\Psi}_{3,4}=&\frac{1}{2\sqrt{M}}\sum_{k=1}^{M}\big[\ket{1_k}_3\ket{1_k}_4\otimes\{\ket{H}_3\ket{V}_4+\ket{H}_4\ket{V}_3\}\otimes\\ &\{\ket{\omega_S}_3\ket{\omega_I}_4+\ket{\omega_S}_3\ket{\omega_I}_4\}\big].\label{eq:9}
\end{aligned}
\end{equation}
The subscripts $3$ and $4$ indicate the two distinct spatial modes while $\ket{1_k}_j$ represents presence of a single photon in the $k$-th temporal mode and the spatial mode $j$ and none in the other temporal modes.  One of the entangled photons is sent towards the possible target while the other is stored. Let the spatial mode associated with the dispatched photon be $3$ and the spatial mode occupied by the stored photon be $4$. The local state of the stored photon is of the form,
\begin{equation}
\sigma=\text{Tr}_3\big(\ket{\Psi}\bra{\Psi}_{3,4}\big)=\frac{I_4}{4M}.
\end{equation}
Where $I_4$ is the 4M dimensional identity operator over the space spanned by the single photon states occupying $M$ distinct temporal modes, with each mode being further indexed by two distinct frequencies and polarizations. In the absence of any object, the returning state is the unpolarized thermal state with mean number of thermal photons per temporal mode ($N_B$) satisfying the condition $MN_B \ll 1$. The mean number of thermal photons depends on the frequency but assuming the values of $\omega_S$ and $\omega_I$ are sufficiently close, we approximate the mean thermal photon number to be the same for both the frequencies. The thermal state, $\rho_t$, under the low noise condition is approximated as \,\cite{lloyd2008enhanced},
\begin{align}
\rho_{t}=&(1-M N_B)\ket{0}\bra{0}_3+\frac{N_B}{4}\sum_{k=1}^M\big[\ket{1_k}\bra{1_k}_3 \otimes\{\ket{H}\bra{H}_3\nonumber\\
&+\ket{V}\bra{V}_3\}\otimes\{\ket{\omega_S}\bra{\omega_S}_3 +\ket{\omega_I}\bra{\omega_I}_3\}\big].\\
=&(1-M N_B)\ket{0}\bra{0}_3+\frac{N_B}{4}I_3.
\end{align}
Here $I_3$, like $I_4$, is an identity operator over the single photon subspace. In a slight abuse of notation, we have indexed the returning state as well as the dispatched single photon state by $3$. In the absence of any object, the global state, $\rho_0$ ( i.e. returning state + stored state) is,
\begin{equation}
\rho_0=\rho_t\otimes\sigma.
\end{equation}
  In the presence of a weakly reflecting object with reflectance $\kappa$, we have,
\begin{equation}
\rho_1=(1-\kappa)\rho_0+\kappa\ket{\Psi}\bra{\Psi}_{3,4} .
\end{equation}
The minimum error probability for distinguishing between $\rho_0$ and $\rho_1$ is bounded by the quantum Chernoff bound\,\cite{calsamiglia2008quantum}. In order to determine QCB, we need to evaluate $\rho_0^{1-s}$ and $\rho_1^s$,
\begin{small}
\begin{align}
\rho_0^{1-s}&=\big((1-MN_B)^{1-s}\ket{0}\bra{0}_3+\Big(\frac{N_B}{4}\Big)^{1-s}I_3\big)\otimes\frac{I_4}{{(4M)}^{1-s}}\\
\rho_1^s&=(1-\kappa)^s(1-MN_B)^s\ket{0}\bra{0}_3\otimes\frac{I_4}{(4M)^s}+\nonumber\\
&(1-\kappa)^s\Big(\frac{N_B}{16M}\Big)^s(I_3\otimes I_4-\ket{\Psi}\bra{\Psi}_{3,4})+\nonumber\\
&((1-\kappa)\frac{N_B}{16M}+\kappa)^s\ket{\Psi}\bra{\Psi}_{3,4}.
\end{align}
\end{small}
In arriving at the preceding two equations, we have used the fact that the vacuum state is orthogonal to the single photon subspace and that $(I_3\otimes I_4-\ket{\Psi}\bra{\Psi}_{3,4})$, in addition to being orthogonal to vacuum state, is also orthogonal to $\ket{\Psi}\bra{\Psi}_{3,4}$. Taking the trace of $\rho_0^{1-s}\rho_1^s$ yields,
\begin{align}
 &Q_{QCB}\nonumber\\
 =&\underset{0\leq s\leq 1}{\text{min}}(1-\kappa)^s \Big(1+\frac{N_B}{16M}\big(-1+(1+\frac{16\kappa M}{(1-\kappa)N_B})^s\big)\Big)\\
=& \underset{0\leq s\leq 1}{\text{min}} \Big(1-\kappa s+\frac{N_B}{16M}\big(-1+(1+\frac{16\kappa M}{N_B})^s\big)\Big)\nonumber\\
& + O(N_B^2,\kappa N_B)
\end{align}
We now focus on the so-called bad regime, which is characterized by $MN_B \ll 1$ and $\kappa \ll N_B/M$. The second condition allows us to approximate $(1+\frac{16\kappa M}{N_B})^s$ by expanding the term up to second order in $\kappa M/N_B$ to give,
\begin{align}
Q_{QCB}&\approx \underset{0\leq s\leq 1}{\text{min}}\bigg(1+s(s-1)\frac{16\kappa^2M}{2N_B}\bigg)\label{eqn:qcb1},\\
Q_{QCB}&\approx 1-\frac{2\kappa^2M}{N_B}.
\end{align}
We see from \eqref{eqn:qcb1} that the minimum occurs at $s=1/2$. Substituting the expression for $Q_{QCB}$ in \eqref{eqn:0} gives an upper bound on the error probability,
\begin{equation}
p_e^{(N)}\leq \frac{1}{2} \Big(1-\frac{2\kappa^2M}{N_B}\Big)^N \approx \frac{1}{2}e^{-2N\kappa^2M/N_B}.
\end{equation}
Comparing with the quantum Chernoff bound for quantum illumination using entangled photon pairs in the same parameter regime \cite{shapiro2009quantum}, we see a gain of a factor of 16 (i.e. 12dB gain) in the probability error exponent for hyperentangled photon pairs. This result can be generalised to hyperentanglement in $f$ degrees of freedom where we obtain a factor of $2^{2f}$ improvement in the exponent, provided each degree assumes only two discrete values. This result anticipates the improvement offered by hyperentanglement even in the high noise regime, which is demonstrated by the performance analysis of the proposed receiver structure in the following sections. For the high noise and low signal intensity regime, we consider the  hyperentangled state obtained after \eqref{eq:beam1} and \eqref{eq:beam2} in the generation procedure, as opposed to the biphoton state considered in low noise regime.
\section{Hyperentanglement-enhanced OPA} 
The phase-sensitive cross correlations of the hyperentangled state are pivotal to the operation of the hyperentaglement-enhanced OPA (as well as the FF-SFG receiver). We explicitly calculate these phase-sensitive cross-correlations.
 \begin{align}
&\langle a_{3,H,S}a_{4,V,I}\rangle  \nonumber\\
&=\bra{\Phi}\Big(\frac{a_{1,H,S}+ia_{2,H,S}}{\sqrt{2}}\Big)\Big(\frac{ia_{1,V,I}+a_{2,V,I}}{\sqrt{2}}\Big)\ket{\Phi}.
\end{align}
 Observing that $\bra{\Phi}a_{1,H,S}a_{2,V,I}\ket{\Phi}$=$\bra{\Phi}a_{2,H,S}a_{1,V,I}\ket{\Phi}=0$, the above equation can be further simplified,
\begin{align}
&\langle a_{3,H,S}a_{4,V,I}\rangle \nonumber\\
&=\frac{i}{2}\bra{\Phi}a_{1,H,S}a_{1,V,I}\ket{\Phi} + \frac{i}{2}\bra{\Phi}a_{2,H,S}a_{2,V,I}\ket{\Phi}\\
&=\frac{i}{4}\sum_{n=1}^\infty\frac{(N_S^\prime)^{n-1/2}}{(N_S^\prime+1)^{n+1/2}}n + \frac{i}{4}\sum_{n=1}^\infty\frac{(N_S^\prime)^{n-1/2}}{(N_S^\prime+1)^{n+1/2}}n\\
&=\frac{i}{2}\sqrt{(N_S^\prime)(N_S^\prime+1)}
\end{align}
We can calculate $\langle a_{3,V,S}a_{4,H,I}\rangle$, $\langle a_{3,H,I}a_{4,V,S}\rangle$ and $\langle a_{3,V,I}a_{4,H,S}\rangle$ along similar lines with the result that all four phase-sensitive cross-correlations possess identical value--$\frac{i}{2}\sqrt{(N_S^\prime)(N_S^\prime+1)}$. If we were to use a separable state instead of the hyperentangled state, then there would be either no initial correlations between the two spatial modes (for a direct
product state) or the initial classical correlations would be much lower than entanglement-induced
correlations with the result that these correlations would deteriorate even further due to thermal
noise. We exploit these correlations using a modified implementation of an OPA based receiver\,\cite{guha2009gaussian}, as demonstrated in the schematic representation, Fig.\,\ref{fig:2}. One of the spatial modes, say $3$, is stored, whereas the other spatial mode is dispatched towards the presumed target. In the absence of any object in the path, the returning mode $a_R$ is simply the thermal bath mode $a_B$. When the object is present, the returning mode is of the form,
 \begin{equation}
  a_R=\sqrt{\kappa}a_4+\sqrt{1-\kappa}a_B.
 \end{equation}
In either case, the stored and the returning mode are subjected to two level splitting, the first based on the polarization and the second based on the frequency, using a polarizing beam splitter and optical grating respectively (See Fig.\,\ref{fig:2}). As the thermal state is completely unpolarized, the thermal photons get divided equally between $a_{R,H}$ and $a_{R,V}$. Moreover, we may assume the thermal photons get distributed roughly equally between the signal and idler frequency modes if the two frequencies have comparable magnitude. At the end of this splitting procedure, each of the four pairs forms the input to  type-II OPA  having gain $G=1+\epsilon^2$, marginally greater than one. Let us consider the first of these parametric amplifiers with output $c$,
\begin{equation}
c=\sqrt{G}a_{3,H,S}+i\sqrt{G-1}a_{4,V,I}^\dagger.
\end{equation}
This output is a thermal state with mean photon number $N_1$ in presence of the object and $N_0$ in its absence,
\begin{align}
 N_0&= G \frac{N_S^\prime}{2} +(G-1)\bigg(\frac{N_B}{4}+1\bigg), \\
N_1&= G \frac{N_S^\prime}{2} + (G-1)\bigg(\frac{N_B}{4}+ \kappa \frac{N_S^\prime}{2} + 1\bigg) \nonumber \\ 
&+ \sqrt{G(G-1)\kappa N_S^\prime(N_S^\prime+1)} .  
\end{align}

Such a distribution of photons\,\cite{guha2009gaussian} at an ideal photocounter will have variance $\sigma_m^2=N_m(N_m+1)$ where $m$ assumes the value of $1$ or $0$. The output of all four OPAs are identical in mean photon number and variance. A common photocounter is used to determine the photon count, $N_{pc}$ over all $N$ transmitted temporal modes and corresponding four amplifier outputs. The common counter needs to be positioned such that the output from each of the four amplifiers arrives at the counter sequentially. Alternately, four identical photocounters can be used for outputs of the four amplifiers. From classical detection theory\,\cite{lehmann2006testing}, it follows that the threshold is $N_{th}=4N(\sigma_0N_1+\sigma_1N_0)/(\sigma_0+\sigma_1)$. When the photon number count $N_{pc}>N_{th}$, we declare target to be present and vice versa. The error probability in detection, $p_{e,H-OPA}^{(N)}$ is,
\begin{equation}
p_{e,H-OPA}^{(N)}=\frac{1}{2}\text{erfc}(2\sqrt{RN})\leq \frac{e^{-4NR}}{4\sqrt{\pi NR}},
\end{equation}
where $R$ is the signal-to-noise ratio, 
\begin{equation}
R=\frac{(N_1-N_0)^2}{2(\sigma_1+\sigma_0)^2}\approx \frac{\kappa N_S}{4N_B} .
\end{equation}
The final approximation is valid in the limit $N_S,\kappa, \epsilon \ll 1$ and $N_B\gg 1$. The signal-to-noise ratio for an ideal photocounter in our setup is half of that of OPA receiver used for TMSV probe. However, we have achieved a four-fold multiplicity in number of readings for every iteration of transmitter pulse, leading to a net 3dB gain in error exponent over lone OPA receiver. In Fig.\ref{fig:1}, we compare the performance of this receiver with earlier receiver models.
\section{Hyperentanglement-enhanced FF-SFG receiver} If we replace each of the four OPA receivers with FF-SFG receivers, the resulting receiver setup outperforms the lone FF-SFG receiver. For a lone SFG receiver used in combination with $N$ iterations of TMSV probe state, the returning mode, $a_R$ and the idler mode,$a_I$ form the two inputs to give an output mode, $b$ whose frequency is sum of the frequencies of the inputs. Due to practical considerations, $K$ feed-forward cycles are carried out and the mean photon number of SFG output at the end of each of the $K$ cycles is determined. For sufficiently large $K$ and parameter regime--$\kappa\ll 1, N_S\ll 1$ and $N_B\gg 1$, the coherent contribution to mean  photon number in the absence of any object is nearly zero while in the presence of an object\,\citep{zhuang2017optimum}, it assumes the form,
\begin{equation}
\sum_{k=1}^K\langle {b^{(k)}}^\dagger b^{(k)}\rangle=N|\langle a_Ra_I\rangle|^2/(1+N_B)\approx \kappa NN_S/N_B.
\end{equation}  
Here, $b^{(k)}$ is the output at the end of $k$th cycle. The quantity in the preceding equation has been shown to be the exponent in error probability bound for FF-SFG receiver\,\cite{zhuang2017optimum}. The error probability bound, thus, has the following form,
\begin{equation}
\sum_{k=1}^K\langle {b^{(k)}}^\dagger b^{(k)}\rangle=N|\langle a_Ra_I\rangle|^2/(1+N_B)\approx \kappa NN_S/N_B. \label{eqn:ffsfg}
\end{equation} 
The hyperentanglement enhanced FF-SFG receiver model consist of four identical FF-SFG receivers which differ solely in their input modes. The input mode pairs for the four receivers following the hyperentanglement based splitting procedure are $\{a_{3,H,S}, a_{R,V,I}\}, \{a_{3,V,S}, a_{R,H,I}\}, \{a_{3,H,I}, a_{R,V,S}\}$ and $\{a_{3,V,I}, a_{R,H,S}\}$. We focus on the first of these receivers with input modes $a_{3,H,S}$ and $a_{R,V,I}$. The error probability exponent for this receiver is calculated by replacing  $|\langle a_Ra_I\rangle|^2$ in \eqref{eqn:ffsfg} with $|\langle a_{3,H,S}, a_{R,V,I}\rangle|^2$ and $N_B$ by $N_B/4$. This is because, the thermal photons are distributed almost equally among the four modes -- $a_{R,V,I}, a_{R,H,I}a_{R,V,S}$ and $a_{R,H,S}$ following the splitting procedure,
\begin{align}
&\langle a_{3,H,S}a_{R,V,I}\rangle\\
&=\langle a_{3,H,S}(\sqrt{\kappa}a_{4,V,I}+\sqrt{1-\kappa}a_{B,V,I})\rangle=\sqrt{\kappa}\langle a_{3,H,S}a_{4,V,I}\rangle\\
&=\frac{i}{2}\sqrt{\kappa(N_S^\prime)(N_S^\prime+1)}=\frac{i}{2\sqrt{2}}\sqrt{\kappa(N_S)(N_S/2+1)}.
\end{align} 
We have used the fact that $N_S^\prime =N_S/2$. Substituting the above result in \eqref{eqn:ffsfg} and approximating the resulting expression under the conditions $N_S \ll 1$ and $N_B\gg 1$ yields,
\begin{align}
\sum_{k=1}^K\langle {b^{(k)}}^\dagger b^{(k)}\rangle=\frac{\kappa NN_S(N_S/2+1)}{8(1+N_B/4)}\approx \kappa NN_S/2N_B
\end{align}
The error probability is, therefore bounded as follows,
\begin{equation}
p_{e,1}^{(N)}\leq e^{-\kappa NN_S/2N_B}.
\end{equation}
The subscript 1 in error probability indicates the first of four FF-SFG receivers. The total error probability is given by,
\begin{equation}
p_{e}^{(N)}=\prod_{i=1}^4 p_{e,i}^{(N)}
\end{equation}
As the phase sensitive correlations appearing in the exponent of error probability for each receiver are identical, it follows that
\begin{equation}
p_{e}^{(N)}=(p_{e,1}^{(N)})^4\leq e^{-2\kappa NN_S/N_B}.
\end{equation}
 We see a factor of 2 (i.e. 3dB) improvement in the exponent over the performance of a lone FF-SFG receiver. Thus, hyperentanglement-enhanced FF-SFG receiver outperforms the optimal quantum illumination receiver by 3dB in error probability exponent.\\
\section{Concluding Remarks} 
In this article, we have shown the benefits of hyperentanglement in low noise as well as high noise regime with a 12dB improvement over QI in low noise regime and two variants of hyperentanglement enhanced receiver in high noise regime matching FF-SFG receiver performance in one instance and surpassing it in the other. The main advantages of the receiver model proposed in this article stem from the distribution of thermal noise photons into four modes following the splitting procedure. Hyperentanglement enables us to achieve such a distribution without nullifying the phase sensitive cross correlations. For the probe state considered, the performance of the proposed receivers is (possibly) sub-optimal and only a complete Chernoff bound computation of the hyperentangled probe states using symplectic decomposition will shed light on the optimal performance\,\cite{pirandola2008computable}. However, it should be noted that despite possibly being suboptimal, hyperentanglement-enhanced FF-SFG receiver outperforms the lone FF-SFG receiver, making it the best receiver model till date. The hyperentanglement based splitting procedure in our model is general enough to allow for other degrees of freedom like time-bin to replace polarization or frequency in order to circumvent any practical issues associated with frequency and polarization hyperentangled states. In fact, if we are able to exploit entanglement in a third degree of freedom we will be able to surpass the performance of the hyperentanglement-enhanced OPA by 3dB. In conclusion, the proposed receiver structure is highly versatile and lends itself to several modifications that need to be further investigated.

\section{Acknowledgement}
 AVP, BS and CMC acknowledge the support of Kishore Vaigyanik Protsahan Yojana (KVPY), Infosys Young Investigator grant and Ramanujan Fellowship, respectively.


\begin{thebibliography}{22}%
\makeatletter
\providecommand \@ifxundefined [1]{%
 \@ifx{#1\undefined}
}%
\providecommand \@ifnum [1]{%
 \ifnum #1\expandafter \@firstoftwo
 \else \expandafter \@secondoftwo
 \fi
}%
\providecommand \@ifx [1]{%
 \ifx #1\expandafter \@firstoftwo
 \else \expandafter \@secondoftwo
 \fi
}%
\providecommand \natexlab [1]{#1}%
\providecommand \enquote  [1]{``#1''}%
\providecommand \bibnamefont  [1]{#1}%
\providecommand \bibfnamefont [1]{#1}%
\providecommand \citenamefont [1]{#1}%
\providecommand \href@noop [0]{\@secondoftwo}%
\providecommand \href [0]{\begingroup \@sanitize@url \@href}%
\providecommand \@href[1]{\@@startlink{#1}\@@href}%
\providecommand \@@href[1]{\endgroup#1\@@endlink}%
\providecommand \@sanitize@url [0]{\catcode `\\12\catcode `\$12\catcode
  `\&12\catcode `\#12\catcode `\^12\catcode `\_12\catcode `\%12\relax}%
\providecommand \@@startlink[1]{}%
\providecommand \@@endlink[0]{}%
\providecommand \url  [0]{\begingroup\@sanitize@url \@url }%
\providecommand \@url [1]{\endgroup\@href {#1}{\urlprefix }}%
\providecommand \urlprefix  [0]{URL }%
\providecommand \Eprint [0]{\href }%
\providecommand \doibase [0]{https://doi.org/}%
\providecommand \selectlanguage [0]{\@gobble}%
\providecommand \bibinfo  [0]{\@secondoftwo}%
\providecommand \bibfield  [0]{\@secondoftwo}%
\providecommand \translation [1]{[#1]}%
\providecommand \BibitemOpen [0]{}%
\providecommand \bibitemStop [0]{}%
\providecommand \bibitemNoStop [0]{.\EOS\space}%
\providecommand \EOS [0]{\spacefactor3000\relax}%
\providecommand \BibitemShut  [1]{\csname bibitem#1\endcsname}%
\let\auto@bib@innerbib\@empty
\bibitem [{\citenamefont {Pirandola}\ \emph {et~al.}(2018)\citenamefont
  {Pirandola}, \citenamefont {Bardhan}, \citenamefont {Gehring}, \citenamefont
  {Weedbrook},\ and\ \citenamefont {Lloyd}}]{pirandola2018advances}%
  \BibitemOpen
  \bibfield  {author} {\bibinfo {author} {\bibfnamefont {S.}~\bibnamefont
  {Pirandola}}, \bibinfo {author} {\bibfnamefont {B.~R.}\ \bibnamefont
  {Bardhan}}, \bibinfo {author} {\bibfnamefont {T.}~\bibnamefont {Gehring}},
  \bibinfo {author} {\bibfnamefont {C.}~\bibnamefont {Weedbrook}},\ and\
  \bibinfo {author} {\bibfnamefont {S.}~\bibnamefont {Lloyd}},\ }\href@noop {}
  {\bibfield  {journal} {\bibinfo  {journal} {Nature Photonics}\ }\textbf
  {\bibinfo {volume} {12}},\ \bibinfo {pages} {724} (\bibinfo {year}
  {2018})}\BibitemShut {NoStop}%
\bibitem [{\citenamefont {Shapiro}(2020)}]{shapiro2020quantum}%
  \BibitemOpen
  \bibfield  {author} {\bibinfo {author} {\bibfnamefont {J.~H.}\ \bibnamefont
  {Shapiro}},\ }\href@noop {} {\bibfield  {journal} {\bibinfo  {journal} {IEEE
  Aerospace and Electronic Systems Magazine}\ }\textbf {\bibinfo {volume}
  {35}},\ \bibinfo {pages} {8} (\bibinfo {year} {2020})}\BibitemShut {NoStop}%
\bibitem [{\citenamefont {Helstrom}\ and\ \citenamefont
  {Helstrom}(1976)}]{helstrom1976quantum}%
  \BibitemOpen
  \bibfield  {author} {\bibinfo {author} {\bibfnamefont {C.~W.}\ \bibnamefont
  {Helstrom}}\ and\ \bibinfo {author} {\bibfnamefont {C.~W.}\ \bibnamefont
  {Helstrom}},\ }\href@noop {} {\emph {\bibinfo {title} {Quantum detection and
  estimation theory}}},\ Vol.~\bibinfo {volume} {3}\ (\bibinfo  {publisher}
  {Academic press New York},\ \bibinfo {year} {1976})\BibitemShut {NoStop}%
\bibitem [{\citenamefont {Audenaert}\ \emph {et~al.}(2007)\citenamefont
  {Audenaert}, \citenamefont {Calsamiglia}, \citenamefont {Munoz-Tapia},
  \citenamefont {Bagan}, \citenamefont {Masanes}, \citenamefont {Acin},\ and\
  \citenamefont {Verstraete}}]{audenaert2007discriminating}%
  \BibitemOpen
  \bibfield  {author} {\bibinfo {author} {\bibfnamefont {K.~M.}\ \bibnamefont
  {Audenaert}}, \bibinfo {author} {\bibfnamefont {J.}~\bibnamefont
  {Calsamiglia}}, \bibinfo {author} {\bibfnamefont {R.}~\bibnamefont
  {Munoz-Tapia}}, \bibinfo {author} {\bibfnamefont {E.}~\bibnamefont {Bagan}},
  \bibinfo {author} {\bibfnamefont {L.}~\bibnamefont {Masanes}}, \bibinfo
  {author} {\bibfnamefont {A.}~\bibnamefont {Acin}},\ and\ \bibinfo {author}
  {\bibfnamefont {F.}~\bibnamefont {Verstraete}},\ }\href@noop {} {\bibfield
  {journal} {\bibinfo  {journal} {Physical review letters}\ }\textbf {\bibinfo
  {volume} {98}},\ \bibinfo {pages} {160501} (\bibinfo {year}
  {2007})}\BibitemShut {NoStop}%
\bibitem [{\citenamefont {Lloyd}(2008)}]{lloyd2008enhanced}%
  \BibitemOpen
  \bibfield  {author} {\bibinfo {author} {\bibfnamefont {S.}~\bibnamefont
  {Lloyd}},\ }\href@noop {} {\bibfield  {journal} {\bibinfo  {journal}
  {Science}\ }\textbf {\bibinfo {volume} {321}},\ \bibinfo {pages} {1463}
  (\bibinfo {year} {2008})}\BibitemShut {NoStop}%
\bibitem [{\citenamefont {Tan}\ \emph {et~al.}(2008)\citenamefont {Tan},
  \citenamefont {Erkmen}, \citenamefont {Giovannetti}, \citenamefont {Guha},
  \citenamefont {Lloyd}, \citenamefont {Maccone}, \citenamefont {Pirandola},\
  and\ \citenamefont {Shapiro}}]{tan2008quantum}%
  \BibitemOpen
  \bibfield  {author} {\bibinfo {author} {\bibfnamefont {S.-H.}\ \bibnamefont
  {Tan}}, \bibinfo {author} {\bibfnamefont {B.~I.}\ \bibnamefont {Erkmen}},
  \bibinfo {author} {\bibfnamefont {V.}~\bibnamefont {Giovannetti}}, \bibinfo
  {author} {\bibfnamefont {S.}~\bibnamefont {Guha}}, \bibinfo {author}
  {\bibfnamefont {S.}~\bibnamefont {Lloyd}}, \bibinfo {author} {\bibfnamefont
  {L.}~\bibnamefont {Maccone}}, \bibinfo {author} {\bibfnamefont
  {S.}~\bibnamefont {Pirandola}},\ and\ \bibinfo {author} {\bibfnamefont
  {J.~H.}\ \bibnamefont {Shapiro}},\ }\href@noop {} {\bibfield  {journal}
  {\bibinfo  {journal} {Physical review letters}\ }\textbf {\bibinfo {volume}
  {101}},\ \bibinfo {pages} {253601} (\bibinfo {year} {2008})}\BibitemShut
  {NoStop}%
\bibitem [{fno()}]{fnote}%
  \BibitemOpen
  \href@noop {} {}\bibinfo {note} {In high noise, low signal intensity and high
  loss regime, the quantum Chernoff bound for coherent states is $e^{-\kappa
  N_S/4N_B}/2$. This bound is staturated using a homodyne detector}\BibitemShut
  {NoStop}%
\bibitem [{\citenamefont {Shapiro}\ and\ \citenamefont
  {Lloyd}(2009)}]{shapiro2009quantum}%
  \BibitemOpen
  \bibfield  {author} {\bibinfo {author} {\bibfnamefont {J.~H.}\ \bibnamefont
  {Shapiro}}\ and\ \bibinfo {author} {\bibfnamefont {S.}~\bibnamefont
  {Lloyd}},\ }\href@noop {} {\bibfield  {journal} {\bibinfo  {journal} {New
  Journal of Physics}\ }\textbf {\bibinfo {volume} {11}},\ \bibinfo {pages}
  {063045} (\bibinfo {year} {2009})}\BibitemShut {NoStop}%
\bibitem [{\citenamefont {Sacchi}(2005)}]{sacchi2005optimal}%
  \BibitemOpen
  \bibfield  {author} {\bibinfo {author} {\bibfnamefont {M.~F.}\ \bibnamefont
  {Sacchi}},\ }\href@noop {} {\bibfield  {journal} {\bibinfo  {journal}
  {Physical Review A}\ }\textbf {\bibinfo {volume} {71}},\ \bibinfo {pages}
  {062340} (\bibinfo {year} {2005})}\BibitemShut {NoStop}%
\bibitem [{\citenamefont {Kwiat}(1997)}]{kwiat1997hyper}%
  \BibitemOpen
  \bibfield  {author} {\bibinfo {author} {\bibfnamefont {P.~G.}\ \bibnamefont
  {Kwiat}},\ }\href@noop {} {\bibfield  {journal} {\bibinfo  {journal} {Journal
  of modern optics}\ }\textbf {\bibinfo {volume} {44}},\ \bibinfo {pages}
  {2173} (\bibinfo {year} {1997})}\BibitemShut {NoStop}%
\bibitem [{\citenamefont {Deng}\ \emph {et~al.}(2017)\citenamefont {Deng},
  \citenamefont {Ren},\ and\ \citenamefont {Li}}]{deng2017quantum}%
  \BibitemOpen
  \bibfield  {author} {\bibinfo {author} {\bibfnamefont {F.-G.}\ \bibnamefont
  {Deng}}, \bibinfo {author} {\bibfnamefont {B.-C.}\ \bibnamefont {Ren}},\ and\
  \bibinfo {author} {\bibfnamefont {X.-H.}\ \bibnamefont {Li}},\ }\href@noop {}
  {\bibfield  {journal} {\bibinfo  {journal} {Science bulletin}\ }\textbf
  {\bibinfo {volume} {62}},\ \bibinfo {pages} {46} (\bibinfo {year}
  {2017})}\BibitemShut {NoStop}%
\bibitem [{\citenamefont {Walborn}\ \emph {et~al.}(2018)\citenamefont
  {Walborn}, \citenamefont {Pimentel}, \citenamefont {Davidovich},\ and\
  \citenamefont {de~Matos~Filho}}]{walborn2018quantum}%
  \BibitemOpen
  \bibfield  {author} {\bibinfo {author} {\bibfnamefont {S.}~\bibnamefont
  {Walborn}}, \bibinfo {author} {\bibfnamefont {A.}~\bibnamefont {Pimentel}},
  \bibinfo {author} {\bibfnamefont {L.}~\bibnamefont {Davidovich}},\ and\
  \bibinfo {author} {\bibfnamefont {R.}~\bibnamefont {de~Matos~Filho}},\
  }\href@noop {} {\bibfield  {journal} {\bibinfo  {journal} {Physical Review
  A}\ }\textbf {\bibinfo {volume} {97}},\ \bibinfo {pages} {010301} (\bibinfo
  {year} {2018})}\BibitemShut {NoStop}%
\bibitem [{\citenamefont {Chen}\ \emph {et~al.}(2020)\citenamefont {Chen},
  \citenamefont {Riazi}, \citenamefont {Zhu},\ and\ \citenamefont
  {Qian}}]{chen2020recovering}%
  \BibitemOpen
  \bibfield  {author} {\bibinfo {author} {\bibfnamefont {C.}~\bibnamefont
  {Chen}}, \bibinfo {author} {\bibfnamefont {A.}~\bibnamefont {Riazi}},
  \bibinfo {author} {\bibfnamefont {E.~Y.}\ \bibnamefont {Zhu}},\ and\ \bibinfo
  {author} {\bibfnamefont {L.}~\bibnamefont {Qian}},\ }\href@noop {} {\bibfield
   {journal} {\bibinfo  {journal} {Physical Review A}\ }\textbf {\bibinfo
  {volume} {101}},\ \bibinfo {pages} {013834} (\bibinfo {year}
  {2020})}\BibitemShut {NoStop}%
\bibitem [{\citenamefont {Guha}\ and\ \citenamefont
  {Erkmen}(2009)}]{guha2009gaussian}%
  \BibitemOpen
  \bibfield  {author} {\bibinfo {author} {\bibfnamefont {S.}~\bibnamefont
  {Guha}}\ and\ \bibinfo {author} {\bibfnamefont {B.~I.}\ \bibnamefont
  {Erkmen}},\ }\href@noop {} {\bibfield  {journal} {\bibinfo  {journal}
  {Physical Review A}\ }\textbf {\bibinfo {volume} {80}},\ \bibinfo {pages}
  {052310} (\bibinfo {year} {2009})}\BibitemShut {NoStop}%
\bibitem [{\citenamefont {Zhang}\ \emph {et~al.}(2015)\citenamefont {Zhang},
  \citenamefont {Mouradian}, \citenamefont {Wong},\ and\ \citenamefont
  {Shapiro}}]{zhang2015entanglement}%
  \BibitemOpen
  \bibfield  {author} {\bibinfo {author} {\bibfnamefont {Z.}~\bibnamefont
  {Zhang}}, \bibinfo {author} {\bibfnamefont {S.}~\bibnamefont {Mouradian}},
  \bibinfo {author} {\bibfnamefont {F.~N.}\ \bibnamefont {Wong}},\ and\
  \bibinfo {author} {\bibfnamefont {J.~H.}\ \bibnamefont {Shapiro}},\
  }\href@noop {} {\bibfield  {journal} {\bibinfo  {journal} {Physical review
  letters}\ }\textbf {\bibinfo {volume} {114}},\ \bibinfo {pages} {110506}
  (\bibinfo {year} {2015})}\BibitemShut {NoStop}%
\bibitem [{\citenamefont {Zhuang}\ \emph {et~al.}(2017)\citenamefont {Zhuang},
  \citenamefont {Zhang},\ and\ \citenamefont {Shapiro}}]{zhuang2017optimum}%
  \BibitemOpen
  \bibfield  {author} {\bibinfo {author} {\bibfnamefont {Q.}~\bibnamefont
  {Zhuang}}, \bibinfo {author} {\bibfnamefont {Z.}~\bibnamefont {Zhang}},\ and\
  \bibinfo {author} {\bibfnamefont {J.~H.}\ \bibnamefont {Shapiro}},\
  }\href@noop {} {\bibfield  {journal} {\bibinfo  {journal} {Physical review
  letters}\ }\textbf {\bibinfo {volume} {118}},\ \bibinfo {pages} {040801}
  (\bibinfo {year} {2017})}\BibitemShut {NoStop}%
\bibitem [{\citenamefont {Barreiro}\ \emph {et~al.}(2005)\citenamefont
  {Barreiro}, \citenamefont {Langford}, \citenamefont {Peters},\ and\
  \citenamefont {Kwiat}}]{barreiro2005generation}%
  \BibitemOpen
  \bibfield  {author} {\bibinfo {author} {\bibfnamefont {J.~T.}\ \bibnamefont
  {Barreiro}}, \bibinfo {author} {\bibfnamefont {N.~K.}\ \bibnamefont
  {Langford}}, \bibinfo {author} {\bibfnamefont {N.~A.}\ \bibnamefont
  {Peters}},\ and\ \bibinfo {author} {\bibfnamefont {P.~G.}\ \bibnamefont
  {Kwiat}},\ }\href@noop {} {\bibfield  {journal} {\bibinfo  {journal}
  {Physical review letters}\ }\textbf {\bibinfo {volume} {95}},\ \bibinfo
  {pages} {260501} (\bibinfo {year} {2005})}\BibitemShut {NoStop}%
\bibitem [{\citenamefont {Kwiat}\ \emph {et~al.}(1999)\citenamefont {Kwiat},
  \citenamefont {Waks}, \citenamefont {White}, \citenamefont {Appelbaum},\ and\
  \citenamefont {Eberhard}}]{kwiat1999ultrabright}%
  \BibitemOpen
  \bibfield  {author} {\bibinfo {author} {\bibfnamefont {P.~G.}\ \bibnamefont
  {Kwiat}}, \bibinfo {author} {\bibfnamefont {E.}~\bibnamefont {Waks}},
  \bibinfo {author} {\bibfnamefont {A.~G.}\ \bibnamefont {White}}, \bibinfo
  {author} {\bibfnamefont {I.}~\bibnamefont {Appelbaum}},\ and\ \bibinfo
  {author} {\bibfnamefont {P.~H.}\ \bibnamefont {Eberhard}},\ }\href@noop {}
  {\bibfield  {journal} {\bibinfo  {journal} {Physical Review A}\ }\textbf
  {\bibinfo {volume} {60}},\ \bibinfo {pages} {R773} (\bibinfo {year}
  {1999})}\BibitemShut {NoStop}%
\bibitem [{foo()}]{footnote}%
  \BibitemOpen
  \href@noop {} {}\bibinfo {note} {The generation of the described states is experimentally challenging but since $N_S^\prime \ll
  1$, the predominant contribution stems from vacuum state and a pair of
  photons. So, the states, under first order approximation, behave as polarization-entangled states which are routinely generated in labs.}\BibitemShut
  {Stop}%
\bibitem [{\citenamefont {Calsamiglia}\ \emph {et~al.}(2008)\citenamefont
  {Calsamiglia}, \citenamefont {Mu{\~n}oz-Tapia}, \citenamefont {Masanes},
  \citenamefont {Acin},\ and\ \citenamefont {Bagan}}]{calsamiglia2008quantum}%
  \BibitemOpen
  \bibfield  {author} {\bibinfo {author} {\bibfnamefont {J.}~\bibnamefont
  {Calsamiglia}}, \bibinfo {author} {\bibfnamefont {R.}~\bibnamefont
  {Mu{\~n}oz-Tapia}}, \bibinfo {author} {\bibfnamefont {L.}~\bibnamefont
  {Masanes}}, \bibinfo {author} {\bibfnamefont {A.}~\bibnamefont {Acin}},\ and\
  \bibinfo {author} {\bibfnamefont {E.}~\bibnamefont {Bagan}},\ }\href@noop {}
  {\bibfield  {journal} {\bibinfo  {journal} {Physical Review A}\ }\textbf
  {\bibinfo {volume} {77}},\ \bibinfo {pages} {032311} (\bibinfo {year}
  {2008})}\BibitemShut {NoStop}%
\bibitem [{\citenamefont {Lehmann}\ and\ \citenamefont
  {Romano}(2006)}]{lehmann2006testing}%
  \BibitemOpen
  \bibfield  {author} {\bibinfo {author} {\bibfnamefont {E.~L.}\ \bibnamefont
  {Lehmann}}\ and\ \bibinfo {author} {\bibfnamefont {J.~P.}\ \bibnamefont
  {Romano}},\ }\href@noop {} {\emph {\bibinfo {title} {Testing statistical
  hypotheses}}}\ (\bibinfo  {publisher} {Springer Science \& Business Media},\
  \bibinfo {year} {2006})\BibitemShut {NoStop}%
\bibitem [{\citenamefont {Pirandola}\ and\ \citenamefont
  {Lloyd}(2008)}]{pirandola2008computable}%
  \BibitemOpen
  \bibfield  {author} {\bibinfo {author} {\bibfnamefont {S.}~\bibnamefont
  {Pirandola}}\ and\ \bibinfo {author} {\bibfnamefont {S.}~\bibnamefont
  {Lloyd}},\ }\href@noop {} {\bibfield  {journal} {\bibinfo  {journal}
  {Physical Review A}\ }\textbf {\bibinfo {volume} {78}},\ \bibinfo {pages}
  {012331} (\bibinfo {year} {2008})}\BibitemShut {NoStop}%
\end{thebibliography}

%

\end{document}